\documentclass[colorlinks=True,citecolor= blue,urlcolor = blue,linkcolor = blue,backref = page]{aa}  

\usepackage[switch, modulo]{lineno}
\usepackage{orcidlink}
\usepackage{tabularx}
\usepackage{tablefootnote}
\usepackage{graphicx}
\usepackage{natbib}
\bibpunct{(}{)}{;}{a}{}{,} 

\usepackage{txfonts}

\usepackage{amsmath}
\usepackage[normalem]{ulem}

\begin{document} 

   \title{UV slopes of star-forming galaxies in strong-lensing fields at the epoch of reionisation with JWST}

    \author{Giordano Felicioni      \inst{1}    \orcidlink{0009-0001-0778-9038}
    \and Maru\v{s}a Brada\v{c}      \inst{1,2}  \orcidlink{0000-0001-5984-0395}
    \and Yoshihisa Asada            \inst{3}    \orcidlink{0000-0003-3983-5438}
    \and Nicholas S. Martis         \inst{1}    \orcidlink{0000-0003-3243-9969}
    \and Ghassan T. E. Sarrouh      \inst{4}    \orcidlink{0000-0001-8830-2166}
    \and Guillaume Desprez          \inst{5}    \orcidlink{0000-0001-8325-1742}
    \and Gregor Rihtar\v{s}i\v{c}   \inst{1}    \orcidlink{0009-0009-4388-898X}
    \and Naadiyah Jagga             \inst{4}    \orcidlink{0009-0009-9848-3074}
    \and Anishya Harshan            \inst{6, 7} \orcidlink{0000-0001-9414-6382}
    \and Jon Jude\v{z}              \inst{1}    \orcidlink{0009-0000-2101-1938}
    \and Chris J. Willott           \inst{8}    \orcidlink{0000-0002-4201-7367}
    \and Roberto Abraham            \inst{3, 9} \orcidlink{0000-0002-4542-921X}
    \and Gabriel Brammer            \inst{10,11}\orcidlink{0000-0003-2680-005X}
    \and Vince Estrada-Carpenter    \inst{12,13}\orcidlink{0000-0001-8489-2349}
    \and Jasleen Matharu            \inst{14}   \orcidlink{0000-0002-7547-3385}
    \and Adam Muzzin                \inst{15}   
    \and Ga\"el Noirot              \inst{16}
    \and Marcin Sawicki             \inst{17}   \orcidlink{0000-0002-7712-7857}
    \and Sunna Withers              \inst{15}   \orcidlink{0009-0000-8716-7695}
    \and Vladan Markov              \inst{1}    \orcidlink{0000-0002-5694-6124}
    \and Rosa M. M\'erida           \inst{17}   \orcidlink{0000-0001-8115-5845}
    \and Vesna Pirc Jev\v{s}enak    \inst{1}    \orcidlink{0000-0002-3364-4502}
    \and Roberta Tripodi            \inst{18}   \orcidlink{0000-0002-9909-3491}
    }
    \institute{University of Ljubljana, Faculty of Mathematics and Physics, Jadranska ulica 19, SI-1000 Ljubljana, Slovenia 
    \and Department of Physics and Astronomy, University of California Davis, 1 Shields Avenue, Davis, CA 95616, USA        
    \and Dunlap Institute for Astronomy and Astrophysics, 50 St. George Street, Toronto, Ontario, M5S 3H4, Canada 
    \and Department of Physics and Astronomy, York University, 4700 Keele St. Toronto, Ontario, M3J 1P3, Canada 
    \and Kapteyn Astronomical Institute, University of Groningen, P.O. Box 800, 9700AV Groningen, The Netherlands 
    \and Kavli Institute for Cosmology, University of Cambridge, Madingley Road, Cambridge, CB3 0HA, United Kingdom 
    \and Cavendish Laboratory - Astrophysics Group, University of Cambridge, 19 JJ Thomson Avenue, Cambridge, CB3 0HE, UK 
    \and National Research Council of Canada, Herzberg Astronomy \& Astrophysics Research Centre, 5071 West Saanich Road, Victoria, BC, V9E 2E7, Canada 
    \and David A. Dunlap Department of Astronomy and Astrophysics, University of Toronto, 50 St. George Street, Toronto, Ontario, M5S 3H4, Canada 
    \and Cosmic Dawn Center (DAWN), Denmark 
    \and Niels Bohr Institute, University of Copenhagen, Jagtvej 128, DK-2200 Copenhagen N, Denmark 
    \and School of Earth and Space Exploration, Arizona State University, Tempe, AZ 85287, USA 
    \and Beus Center for Cosmic Foundations, Arizona State University, Tempe, AZ 85287, USA 
    \and Max-Planck-Institut f\"ur Astronomie, K\"onigstuhl 17, D-69117 Heidelberg, Germany 
    \and Department of Physics and Astronomy, York University, 4700 Keele St. Toronto, Ontario, M3J 1P3, Canada 
    \and Space Telescope Science Institute, 3700 San Martin Drive, Baltimore, Maryland 21218, USA 
    \and Department of Astronomy and Physics and Institute for Computational Astrophysics, Saint Mary's University, 923 Robie Street, Halifax, Nova Scotia B3H 3C3, Canada 
    \and INAF - Osservatorio Astronomico di Roma, Via Frascati 33, Monte Porzio Catone, 00078, Italy 
    }           
\titlerunning{UV Slopes in star-forming galaxies at the epoch of reionisation with JWST}             
\authorrunning{Felicioni et al.}

  \abstract{UV slopes ($\beta$) are a powerful diagnostic for galaxies at the epoch of reionisation, tracing star formation, interstellar medium ionisation, and the escape fraction $f_{esc}$ of ionising photons. Studies at low and intermediate $z$ suggested a gradual reddening of the UV slope with time and steeper slopes for fainter galaxies. The James Webb Space Telescope (\textit{JWST}) enabled measurements of the UV slope for statistically significant samples of galaxies deep into the epoch of reionisation up to redshift $z\gtrsim10$ and for increasingly faint galaxies. Recent studies of $\beta$ with JWST at $z>7$ revealed a flattening of the trend of the UV slope with respect to redshift and UV magnitude.}
  {We measure $\beta$ at $z>7.5$ using the strong lensing around massive galaxy clusters to observe high-redshift and faint galaxies. The low-brightness regime is of particular interest for reionisation, as most of the recent models of this process posit that numerous faint galaxies are the prime drivers of reionisation}
  {We used NIRCam and NIRSpec observations from CANUCS, Technicolor, JUMPS, Silver Bullet, UNCOVER, and MEGASCIENCE in seven strong-lensing fields in order to find galaxies down to $M_{UV}\sim-16$ and $7.5<z\lesssim12.5$. We measured their $\beta$ with a forward-modelling procedure. We estimated the $f_{esc}$ of a subsample with emission line data using a relation with UV slope, galaxy size, and H$\beta$ equivalent width calibrated from a low-redshift sample.}
  {We found 378 visually confirmed galaxies, including 43 galaxies with a spectrum. We obtained an average UV slope $<\beta>=-2.3\pm0.4$, an average redshift $<z>=8.5\pm1.0$, and an average AB magnitude $M_{UV}=-18\pm1$. We found no significant evolution of $\beta$ in our redshift range, suggesting a flattening of the $\beta-z$  trend above $z\sim7.5$. We found a weak trend between $\beta$ and $M_{UV}$. }
  {The flat trend of $\beta$ at $z>7.5$ indicates a  similarity in average galaxy properties between $300$ and $600 Myr$ after the Big Bang. The weak trend between $\beta$ and $M_{UV}$ suggests an analogous composition for the interstellar medium of low- and high-mass galaxies, likely due to a lack of time to build up a dust reservoir.}

\maketitle

\section{Introduction}
The UV continuum emission redwards of Lyman-$\alpha$ in galaxies is a powerful diagnostic of different physical properties of a galaxy. It is affected by the direct emission of stars, stellar ages and metallicities, and by the nebular emission of gas and absorption by dust. It is characterised as a power law $F_{\lambda}\sim \lambda^{\beta}$, where $F_\lambda$ is the wavelength-specific flux density, and $\beta$ is the characteristic index of the slope. This quantity provides insights into the stellar populations in a galaxy and on its gas and dust content \citep{calzetti_dust_1994}. Steeper UV slopes (lower $\beta$) have been associated with a higher number of young bright stars and a higher proportion of ionising photon emission. In contrast, heated nebular gas emits light towards longer wavelengths, tending to soften the overall UV slope. \par

Early studies on the evolution of the UV slope were conducted using data from the Hubble Space Telescope (HST), investigating the relation between $\beta_{UV}$ and other physical properties. \citet{bouwens_very_2010} and \citet{bouwens_uv-continuum_2014} measured the slopes of over 4000 galaxies at $4<z<8$ and reported that galaxies at earlier cosmic times have steeper UV slopes on average, suggesting that as star formation continues, the dust reservoir builds up, softening the overall UV radiation. They also found that fainter galaxies are bluer on average, which can be understood to mean that fainter galaxies show a weaker attenuation in the UV in the investigated redshift range. On the other hand, \citet{bolamperti_uv-continuum_2023} used HST data to study individual clumps and reported that clumps also follow the same negative $\beta-z$ and $\beta-M_{UV}$ relations. In addition, the slopes of these individual clumps were bluer than the slope of their host galaxy as a whole. 
\par
Since the launch of the James Webb Space Telescope (JWST), it has been possible to measure the UV slope for statistically significant samples of galaxies at redshift $z\gtrsim10$, with individual galaxies reaching $z_{phot}\sim15$ in the latest studies (e.g. \citealp{asada_improving_2025}), and high-redshift galaxy candidates up to $z\sim25$ \citep{perez-gonzalez_rise_2025, whitler_z_2025}. New data from observation programmes in the first years of JWST allowed us to extend these studies well into the epoch of reionisation (EoR). The surveys Cosmic Evolution Early Release Science Survey (CEERS; \citealp{finkelstein_ceers_2023}), JWST Advanced Deep Extragalactic Survey (JADES; \citealp{eisenstein_overview_2023}), Prime Extragalactic Areas for Reionization Science (PEARLS; \citealp{windhorst_jwst_2023}), Grism Lens-Amplified Survey from
Space (GLASS; \citealp{treu_glass-jwst_2022}), the Next Generation Deep Extragalactic Exploratory Public (NGDEEP) survey \citep{bagley_next_2024}, and Ultradeep NIRSpec and NIRCam ObserVations before the Epoch of Reionization (UNCOVER; \citealp{bezanson_jwst_2024}) are only some of the first surveys that continue to provide data for studies at the EoR. Several of these studies confirmed the trend that fainter galaxies have bluer slopes. The first results included \citet{nanayakkara_early_2023} with photometry from GLASS, \citet{topping_uv_2023} with JADES, \citet{cullen_ultraviolet_2024} with a wide sample of photometric data including GLASS, CEERS, JADES, NGDEEP, and UNCOVER, \citet{austin_epochs_2025} with photometry from the EPOCHS photometric sample \citep{conselice_epochs_2025}, and \citet{dottorini_evolution_2024} with spectroscopy from CEERS and JADES. They similarly reported steeper UV slopes at increasing redshifts for galaxies of similar brightness. In contrast, \citet{saxena_hitting_2024}, using spectral data from JADES and the JADES Origin Field (JOF), found increasing reddening in galaxies at $z>9.5$ and a flattening of the $\beta-M_{UV}$ trend in galaxies at $z\gtrsim8$.\par
The UV slope proves to be a powerful diagnostic for reionisation, and it was proposed as an indicator of a high escape fraction ($f_{esc}$) of Lyman-continuum (LyC) photons (\citealp{zackrisson_spectral_2013, chisholm_far-ultraviolet_2022, flury_low-redshift_2022, choustikov_physics_2024}). Several recent studies with JWST observations have found evidence of galaxies with extremely blue UV slopes with $\beta<-2.6$ (e.g. \citealp{topping_uv_2023, cullen_ultraviolet_2024, saxena_hitting_2024, austin_epochs_2025}). This condition is thought to indicate the leakage of LyC radiation (e.g. \citealp{topping_uv_2023}). At low redshift, where direct measurements of $f_{esc}$ are possible, it is possible to find the relation between the two. The HST Low-redshift Lyman-Continuum Survey (LzLCS) (\citealp{flury_low-redshift_2022}) provides a good testing ground for this. \citet{chisholm_far-ultraviolet_2022} found an inverse correlation between $\beta_{UV}$ and $f_{esc}$ and applied it to high-redshift galaxies to determine their escape fraction, while \citet{flury_low-redshift_2022} also discovered that a combination of diagnostics, including $\beta$, the $[OIII]/[OII]$ ratio O32, and the equivalent width of the $H\beta$ line EW($H\beta$), are better indicators of strong LyC leakage. \citet{mascia_closing_2023, mascia_new_2024} concluded that $\beta$ slopes by themselves are not a reliable estimator, finding a relation between measured $f_{esc}$ and diagnostics available at high reshift, including $\beta$, the galaxy half-light radius $r_e$, and emission line diagnostics. They applied the correlation they found in order to estimate $f_{esc}$ for a sample of $4.5 \leq z\leq8$ galaxies with available spectroscopy.\par
All of the mentioned JWST studies are affected by the fact that they relied on either purely spectroscopic observations of individual high-redshift galaxies, limiting the statistics to galaxies selected for observation, or on a wide sample of photometric observations with the Near Infrared Camera (NIRCam) leveraging mostly broad-band filters and only a few medium-band filters, limiting the accuracy of the measurement of $\beta$.\\
Strongly lensed cluster fields offer a great opportunity for probing the Universe at high redshift, as the magnification from these fields would allow us to detect lower-luminosity galaxies, which are thought to be the source of the majority of ionising photons for reionisation. Several of these fields have already been probed by JWST surveys. Of these, the Canadian NIRISS Unbiased Cluster Survey (CANUCS) \citep{willott_near-infrared_2022, sarrouh_canucstechnicolor_2026} has, with the JWST instruments NIRCam, Near Infrared Spectrograph (NIRSpec), and Near Infrared Imager and Slitless Spectrograph (NIRISS), observed galaxies in five different lensed fields, allowing us to observe galaxies at very high redshifts (up to $z\sim13$). This survey was subsequently integrated with Technicolor (PI: Muzzin) and JUMPS (co-PIs: Withers, Muzzin), two follow-up surveys that target three of the five CANUCS fields: Abell 370, MACS0416, and MACS1149. Other surveys probing strongly lensed field are the \textit{Silver Bullet for Dark Matter} (PIs Bradač, Rihtarščič, \citealp{rihtarsic_mapping_2026}), targeting the Bullet cluster field, and UNCOVER, targeting Abell 2744 \citet{bezanson_jwst_2024, suess_medium_2024}.\par
This paper is organised as follows. In \autoref{sec:data} we present the data we used. In \autoref{sec:methodology} we show the procedure for selecting a reliable high-redshift sample and explain the method for measuring the UV slopes. We present our results in \autoref{sec:results}, and we estimate the LyC escape fraction for a subset of our sample. We analyse the results and the evolution of $\beta$ with redshift and with UV magnitude in \autoref{sec:discussion}.  Finally, we summarise our conclusions in \autoref{sec:summary}. Throughout the paper, we assume a $\Lambda$CDM cosmology with the latest results from Planck \citep{aghanim_planck_2020}.

\section{Data} \label{sec:data}
\paragraph{CANUCS - Technicolor - JUMPS}
CANUCS is a JWST Cycle 1 Guaranteed Time Observations (GTO) galaxy survey (PID: 1208; PI: Willott; \citealp{willott_near-infrared_2022}) targeting five galaxy cluster fields: Abell 370, MACS0417, MACS0416, MACS1423, and MACS1149. A massive galaxy cluster at the primary pointing of each cluster field causes the objects behind it to be strongly lensed and magnified. The central region (CLU pointing) of each field has been observed with the instruments NIRCam (in the imaging mode) and NIRISS (imaging and wide-field slitless spectroscopy), and two flanking fields exist for each field: one field that is observed with NIRISS while NIRCam was being pointed at the central CLU pointing, NSF (NIRISS flanking field), and the other field targeted with NIRCam while NIRISS was pointed at the cluster, NCF (NIRCam flanking field). Subsequent observations from \textit{JWST in Technicolor} Cycle 2 programme (PID: 3362; PI: Muzzin) and JWST Ultimate Medium-band Photometric Survey (JUMPS) Cycle 3 programme (PID: 5890 co-PIs: Withers, Muzzin) added medium-band filters for the NCF and the CLU pointings, respectively, of the Abell 370, MACS0416 and MACS1149 fields. Technicolor completes the full-suit of medium- and broad-band NIRCam filters for the NCF pointing, while JUMPS adds observations in the F360M, F430M, F460M, and F480M NIRCam filters for the CLU pointing.\\
Follow-up NIRSpec prism observations with the Micro-shutter Assembly (MSA) were carried out for selected objects in the CLU fields. These objects were selected for individual science cases by the CANUCS team (see also \citealp[\textpilcrow 2.3]{sarrouh_canucstechnicolor_2026}).\par
\paragraph{Bullet cluster}
We also included Bullet cluster data taken with NIRCam in imaging mode and with NIRSpec in multi-object spectroscopy (MOS) mode, observed as part of the GO Program \textit{Silver Bullet for Dark Matter} (PID: 4598; co-PIs: Bradač, Rihtaršič, Sawicki). The NIRCam filters included for this field are F090W, F115W, F150W, F200W, F277W, F356W, F410M, and F444W.\par
\paragraph{GLASS - UNCOVER - MEGASCIENCE}
Finally, we included NIRCam broad- and medium-band observation of the Abell 2744 cluster field from GLASS-JWST-ERS (PID: 1324 PI: Tommaso Treu \citealp{treu_glass-jwst_2022}, and PID: 2756 PI: Wenlai Chen), UNCOVER (PID: 2561; PIs: Ivo Labbe and Rachel Bezanson \citealp{bezanson_jwst_2024}), and MEGASCIENCE (PID: 4111; PI: Wren Suess \citealp{suess_medium_2024}). NIRCam filters covered by the survey are F070W, F090W, F115W, F140M, F150W, F182M, F162M, F200W, F210M, F250M, F277W, F300M, F335M, F356W, F360M, F410M, F430M, F444W, F460M, and F480M. However, not the entire field is covered by every filter, and only the F070W, F090W, F115W, F150W, F200W, F277W, F356, F444W, and F460M filters are available for every object, while the coverage by other filters depends on the galaxy position in the field.
\subsection{Photometric data}
We used NIRCam photometry to calculate the UV slopes of high-redshift $z>7.5$ galaxies in the CLU and NCF fields. NIRISS observations also exist as deep as those conducted with NIRCam, but we chose not to include them as their filters coverage overlaps with that of NIRCam and would have been redundant. We used photometry measured within a circular aperture of $0.3\arcsec$  ($0.32\arcsec$ for the UNCOVER sample) due to the relatively small angular size of most high-redshift sources of interest.\par
For the CANUCS cluster fields, the CLU pointings were mainly observed through all the broad-band filters and the four redder medium-band filters, while all of the NCF pointings were observed through medium- and broad-band filters (see Table \ref{tab:filters}), with the exception of the MACS1149 NCF pointing, which lacks observations in the F162M and F250M filters. The full description of the NIRCam imaging and its processing can be found in the CANUCS Data Release 1 paper by \citet{sarrouh_canucstechnicolor_2026}.\par
Additionally, HST photometry from observations with the Advanced Camera for Surveys (ACS) or the Wide Field Camera 3 (WFC3) for the NCF pointing of MACS0417 and MACS1423 was included to aid in the selection and visual inspection of the sample. The filters we included are F435W, F606W, and F814W (F438W and F606W for fields observed with WFC3).\par
For the Bullet cluster field, we have NIRCam photometry in most broad-band filters and in the F410M medium-band filter.
For Abell 2744, the GLASS, UNCOVER, and MEGASCIENCE surveys provided observations in the full suite of medium- and broad-band filters throughout most of the field.\\

\begin{table}
    \caption{Exposure time for the NIRCam fillters (in ks) in the surveys}
    \label{tab:filters}
    \centering
    \begin{tabular}{ccccc}
    \hline\hline
    Filter    & CLU+   & NCF                  & NCF     & UNCOVER\\
              & Bullet & (HFF)                & (others)\\
    \hline
    F070W     &     - &$10.3$                 &-        &$8.3$\\
    F090W     &$6.4$  &$10.3$                 & $10.3$  &$16.6$\\
    F115W     &$6.4$  &$10.3$                 & $10.3$  &$21.6$\\
    F140M     &    -  &$5.7$                  & $5.7$   &$8.3$\\
    F150W     &$6.4$  &$10.3$                 & $10.3$  &$21.6$\\
    F162M     &    -  &\ $5.7^*$              & $5.7$   &$8.3$\\
    F182M     &    -  &$10.3$                 & $10.3$  &$8.3$\\
    F200W     &$6.4$  &$10$                   &-        &$13.3$\\
    F210M     &    -  &$10.3$                 & $10.3$  &$8.3$\\
    F250M     &    -  &\ $5.7^*$              &$5.7$    &$8.3$\\
    F277M     &$6.4$  &$10.3$                 & $10.3$  &$13.3$\\
    F300M     &    -  &$5.7$                  &$5.7$    &$8.3$\\
    F335M     &    -  &$10.3$                 &$10.3$   &$8.3$\\
    F356W     &$6.4$  &$10$                   &-        &$13.3$\\
    F360M     &\ $9.3^\dagger$  &$10.3$       &$10.3$   &$8.3$\\
    F410M     &$6.4$  &$10.3$                 &$10.3$   &$13.3$\\
    F430M     &\ $9.3^\dagger$       &$10$    & -       &$8.3$\\
    F444W     &$6.4$  &$10.3$                 &$10.3$   &$16.6$\\
    F460M     &\ $9.3^\dagger$      &$10$     & -       &$8.3$\\
    F480M     &\ $9.3^\dagger$      &$10$     & -       &$8.3$\\
    \hline
    \end{tabular}    
    \tablefoot{All NIRCam filters we used and the pointings with their filters, with their respective exposure time in $ks$. CLU and NCF refer to the main and the flanking position of NIRCam in the fields observed by CANUCS. HFF refers to the Hubble Frontier Field cluster fields: Abell 370, MACS1149, and MACS0416. For details about the limiting magnitude in each pointing, see \citealp{sarrouh_canucstechnicolor_2026, bezanson_jwst_2024, suess_medium_2024}\\
    * denotes a filter missing from MACS1149 NCF pointing.\par
    $\dagger$ Only for CANUCS cluster fields.\\
    }

\end{table}

\subsection{Spectroscopic data}
For selected objects in the CLU pointings of CANUCS fields, in the Bullet cluster field, and in Abell 2744, we also have NIRSpec low-resolution prism observations with the MSA, with a typical exposure time of $\sim$2900 s. The raw data were processed with a combination of the standard STScI pipeline and the \verb|MSAEXP| package \citep{brammer_gbrammermsaexp_2022}, including path-loss correction calibrations, in order to obtain 2D spectrum for each source, and 1D spectra were obtained following \citet{horne_optimal_1986}. A full description of the spectroscopy processing is given in \citet{desprez_cdm_2024} and \citet{sarrouh_canucstechnicolor_2026}. For Abell 2744, we used the spectral data from \citet{price_uncover_2025}, available at the DAWN JWST Archive.
\section{Method}\label{sec:methodology}
\subsection{Sample selection}
In order to assemble a reliable high-redshift sample, we obtained a photometric estimate of the redshift with the EAzY SED-fitting code \citep{brammer_eazy_2008} for all of the objects in the fields, using updated models accounting for intergalactic medium (IGM) and circumgalactic medium (CGM) absorption as described in \citet{asada_improving_2025}. \\
A list of high-redshift candidate objects was obtained by applying these criteria, following the same selection procedure as in \citet{willott_steep_2024},
\begin{itemize}
    \item $S/N < 2$ for the F090W filter 
    \item $S/N > 8$ for the F277W filter
    \item $z_{EAzY,ML} > 7.5$
    \item $z_{EAzY,16\%} > 6$,
\end{itemize}
where $z_{EAzY,ML}$ and $z_{EAzY,16\%}$ are the maximum likelihood value and the 16th percentile of the probability density function (PDF) for the redshift parameter in the EAzY SED-fitting procedure, respectively. The first criterion ensured that our objects were undetected blueward of Ly$\alpha$ if they are at $z>7.5$. The F277W criterion constrained our sample to well-detected objects,to avoid objects with poorly determined photometry.  The latter criterion was included to avoid possible interlopers with poorly constrained redshift.\par
\begin{figure}
    \centering
    \includegraphics[width=1\linewidth]{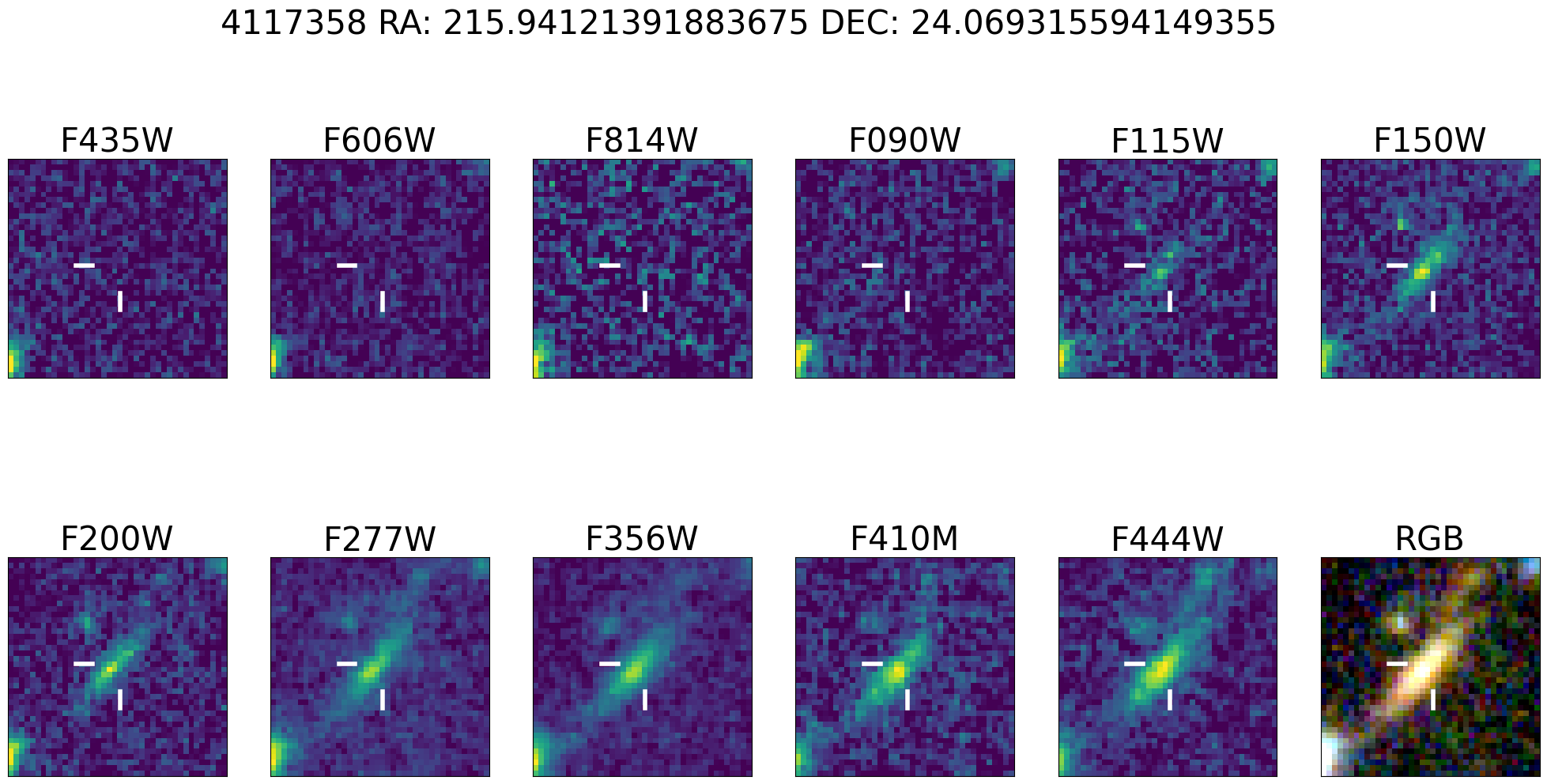}
    \includegraphics[width=1\linewidth]{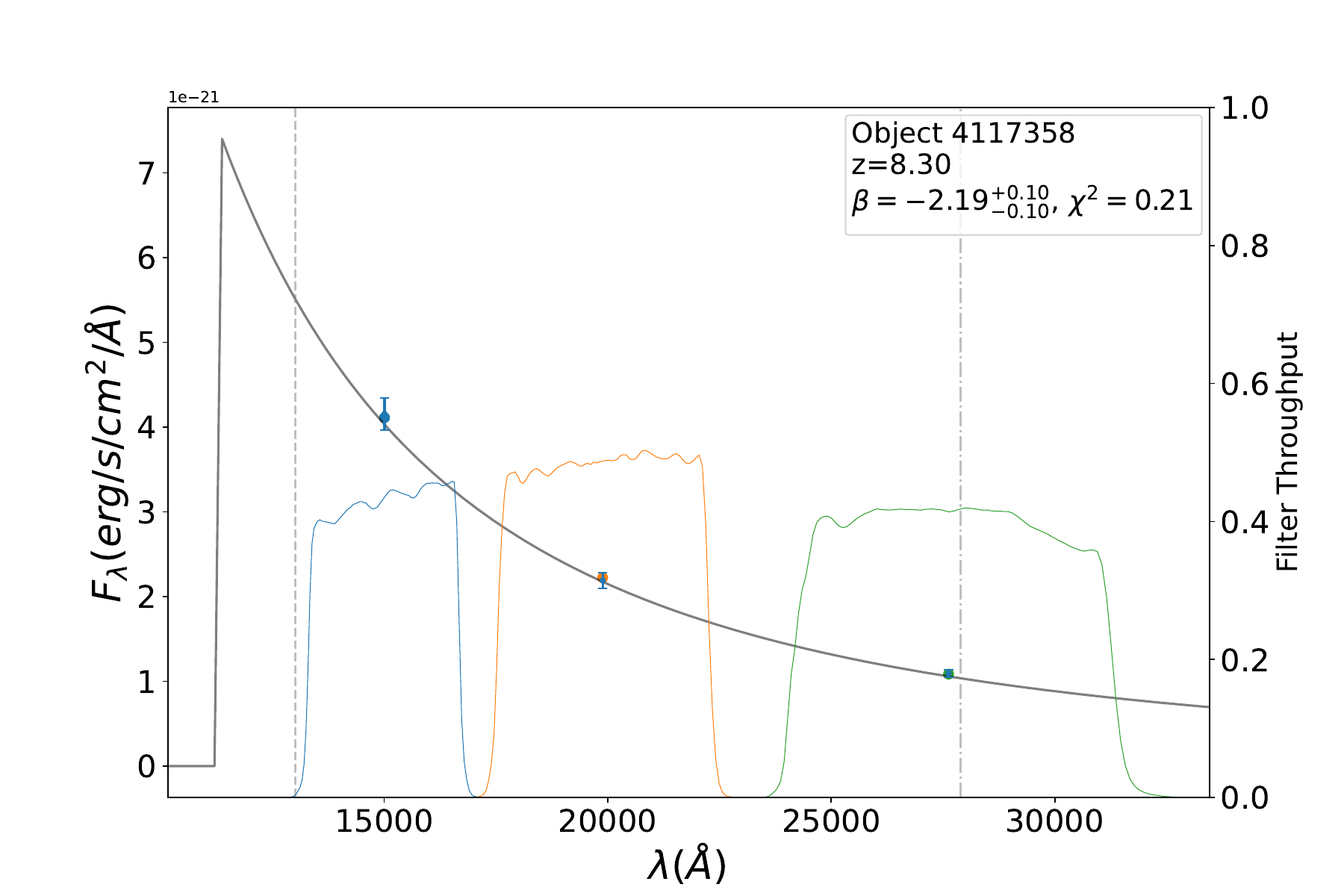}
    \includegraphics[width=1\linewidth]{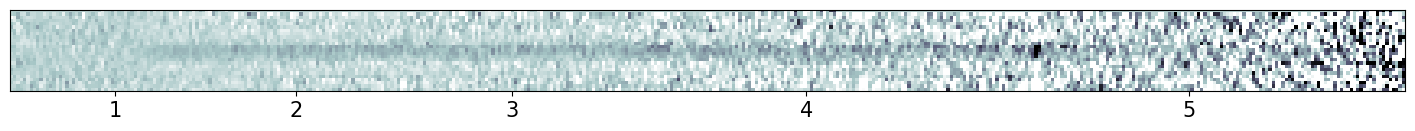}
    \includegraphics[width=1\linewidth]{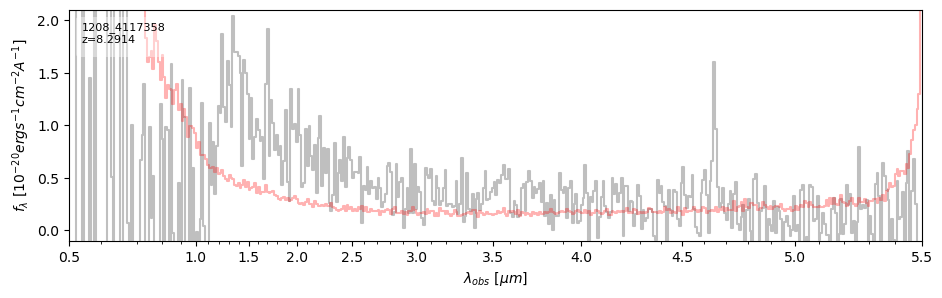}
    \caption{\textit{Top panel}: NIRCam and ACS images of an example object (CANUCS ID 4117358) in the MACS1423 cluster in the CLU pointing in all of the filters available for this pointing.\\ \textit{Middle}: NIRCam photometry overlaid with the throughput of each NIRCam filter, with the fitted $\beta$ slope in yellow. The JWST filter throughput functions are also overlaid in different colours. The vertical dashed (dotted) lines represent the lower (upper) limit for the filters to be used. A hard limit for the lower-wavelength threshold, and a soft limit, only limiting the filter pivot wavelength, for the upper threshold, were applied. In this case, the F150W, F200W, and F277W filters were used to fit the UV slope.\\ \textit{Bottom panel:} 2D and 1D spectrum for the same object. In the 1D spectrum, the grey line represents the signal, and the red line represents the noise at each wavelength of the spectrum. NIRSpec data for realising this graph are now publicly available on the DAWN JWST Archive (DJA) at \texttt{https://dawn-cph.github.io/dja/}}
    \label{fig:placeholder}
\end{figure}

The selected objects were then inspected visually to ensure a non-detection in all filters bluer than the Ly$\alpha$ of all objects, including the HST filters, and to avoid possible interlopers, spurious objects, or image glitches that might have passed the automatic selection (see Figure \ref{fig:placeholder} for an example of an object in the CLU pointing, with available NIRCam broad-band photometry and a NIRSpec prism spectrum). \par

\begin{figure}
    \centering
    \includegraphics[width=\linewidth]{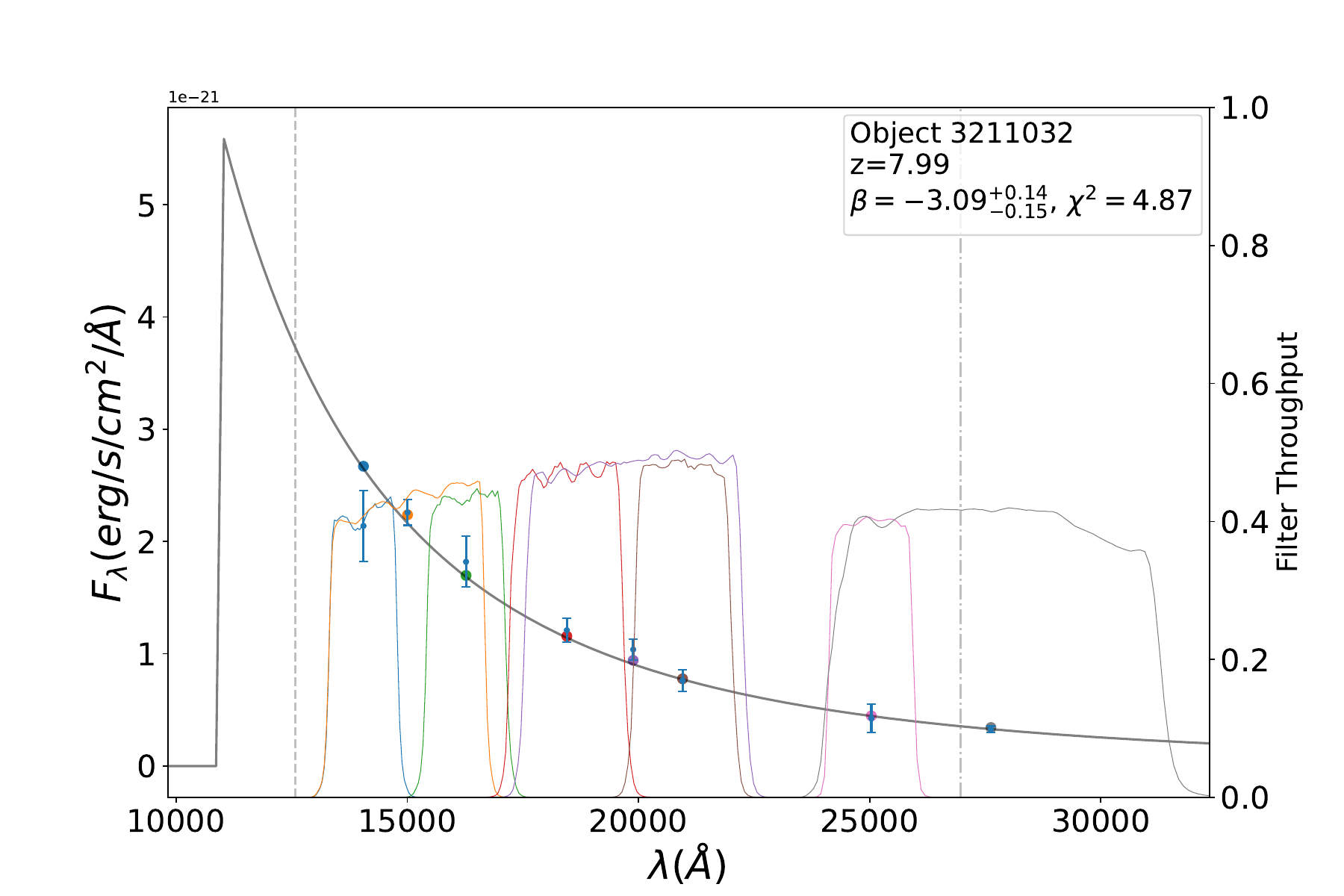}
    \caption{Example of a galaxy at $z=7.87$ from the NCF pointing of the MACS0416 field, whose $\beta$ was calculated from eight NIRCam filters: F140M, F150W, F162M, F182M, F200W, F210M, and F250M}
    \label{fig:8_filters}
\end{figure}

A total of 378 objects were included in our final high-redshift selection. Of these, 212 are part of the five main CANUCS clusters, 36 are in the Bullet cluster, and 130 are from the UNCOVER survey, whose survey area is about three times larger than each of the ten individual CANUCS pointings. Of the objects in the CANUCS clusters, 36 also have NIRSpec observations, and for 24 of these, we used the spectra to compute $\beta$. The $M_{UV}$ of two CANUCS cannot be obtained because photometry in a key filter is lacking, and two further objects are too faint to compute their $r_e$.\par
We also compared the CANUCS objects we found and the objects found using the same criteria in \citet{willott_steep_2024}. We included 57 new objects, and 54 objects were left out in our work because we used a new version of the CANUCS photometric catalog after follow-up observations and updates in the data-processing pipeline: 14 objects in the newer catalog were absent from all in the older versions of the catalog, while 13 objects in the older catalog are not included in the newer version. In both cases, this is due to a combination of factors, such as an object being considered part of a larger nearby object in either of the two versions of the catalog.
In other cases, updated photometry processing resulted in a slightly different S/N in the F090W and F277W filters for borderline cases, and additional filters and updated galaxy templates for fitting their spectral energy distribution (SED) resulted in different values for $z_{_{ML}}$ and $z_{16\%}$ and in a different set of objects to be sampled.\par
\subsection{Spectroscopic sample}
For selected objects of our final photometric sample, we also have NIRSpec observations. For these objects, the spectroscopic redshift $\text{z}_{\text{spec}}$ was used in place of $\text{z}_\text{EAzY,ML}$ when we identified $z>7.5$ galaxies, and the criterion $\text{z}_\text{EAzY,16\%}>6.0$ was ignored altogether, as the spectroscopic redshift is accurate enough to not require a p(z) cut. While we did not carry out a wholesale comparison between spectroscopic and photometric redshift, the following resulted for objects in the CANUCS survey: 44 sources have spectra with a confirmed $z_{spec}>7.5$; of these, 37 are included in our final high-redshift sample. The remaining 7 sources  lack photometry because they are considered part of wider objects. We decided to exclude them as we would lack the NIRCam photometry with which to carry out a photometric correction in order to compute their spectral UV slope. Only 24 of these have a spectrum with high enough signal-to-noise ratio (S/N>1.5) in the wavelength range for measuring their UV slope. We still used the $z_{spec}$ value of the objects we excluded from the spectroscopic sample as the redshift value when measuring $\beta$ from their photometry.
For two of the objects with $z_{spec}$, their $z_{phot}$ alone would have excluded them from the final sample. This suggests that our final sample might be missing 2 for every 35 photometrically selected objects, or $\sim6\%$; the rest of our sample (175 objects selected from photometry alone) would then potentially lack $\sim10$ objects close to the redshift threshold of $z=7.5$.
For the Bullet cluster and Abell 2744 samples, 6 and 13 objects, respectively, have spectroscopy, giving us a total of 43 spectra in all fields.\\

\subsection{Measuring the UV slopes}\label{sec:slopes}
There are different methods for measuring the UV slope of a galaxy. One approach is to fit the emission line-free \citet{calzetti_dust_1994} windows with the galaxy SED obtained from photometric fits, for example using the EAzY code. This would prevent us from finding extremely blue slopes, however, because the SED templates do not include those with $\beta < -3.0$. We decided to rely on a direct power-law fit ($f_\lambda\propto\lambda^\beta$) of the available medium- and wide-band photometry.
For example, \citet{morales_testing_2025} argued in favour of using a direct fit of the photometry for calculating the UV slope, as it recovers the $\beta$ value found from spectroscopy more accurately than other methods, such as SED fitting or single-colour fitting.\par
\subsubsection{Photometric UV slopes}
For each object, we selected the JWST filters with rest-frame wavelengths (at a throughput of more than half its maximum) longer than  $1400\AA$ in order to avoid any possible contamination from the Ly$\alpha$ line, the Ly$\alpha$ damping wing, and the contribution of two-photon emission in the nebular continuum \citep{cameron_nebular_2024}, which peaks at around 1340$\AA$. We used photometric redshifts $z_{phot}$ obtained from EAzY or, where available, its spectroscopic redshift $z_{spec}$.\par  For the long-wavelength boundary of each object, we used a softer threshold by taking filters with a pivot wavelength lower than $3200 \AA$, following the method used in \citet{bolamperti_uv-continuum_2023}. These thresholds leave two to three filters, depending on the redshift of each object, for the measurement of the UV slope for galaxies in the CLU pointings of NIRCam, where only wide-band filters are available, but up to eight wide- and medium-band filters in the NCF pointings (e.g. Fig. \ref{fig:8_filters} for a galaxy with $\beta$ computed from eight NIRCam filters). We estimated the UV slope of each galaxy with a Markov chain Monte Carlo (MCMC) fit that convolved the model UV slope $F_\lambda=A\cdot\lambda^\beta$ with the response function of each filter and compared the forward-modelled photometry with that measured in each filter, with $A$ and $\beta$ as the free parameters to constrain. In order to propagate the redshift uncertainty, we explored the photometric redshift range up to $3\sigma$ from the central value obtained from EAzY, and we divided that redshift range into slices, each with a different set of photometric filters allowed by the criteria described above. We obtained a posterior distribution of the $\beta$ parameter in each of these slices. We computed an average posterior distribution on $\beta$ for all redshift slices by combining the posterior in each of them, weighted by the relative weight of each slice by integrating the $p(z)$ from EAzY. We took the median, 16\% and 84\% of the weighted average posterior distribution of each fit as the central value and the lower and upper confidence levels, respectively. \par

\subsubsection{Spectroscopic UV slopes}
Our spectroscopic sample consists of 43 sources in all fields: 24 in the CANUCS fields, 6 in the Bullet cluster field, and 13 in Abell 2744.\par 
We first adjusted each of the 45 spectra to the photometry of each object. We obtained the spectrum-to-flux ratio by convolving the spectrum with the throuput function of each NIRCam filter and comparing it with the measured NIRCam photometry. We fitted a second-degree Chebyshev polynomial to the spectrum-to-flux ratio points at the pivot wavelength of each filter, and we finally corrected the spectrum with this polynomial, so that the convolution of this corrected spectrum with the NIRCam filters was equal to the photometry measured in each filter. We finally estimated the spectroscopic UV-continuum slope index $\beta_{spec}$ using their photometry-corrected spectrum by fitting a power law through the Calzetti UV-continuum windows to avoid contamination from emission lines and dust absorption. \par
\begin{figure}
    \centering
    \includegraphics[width=1\linewidth]{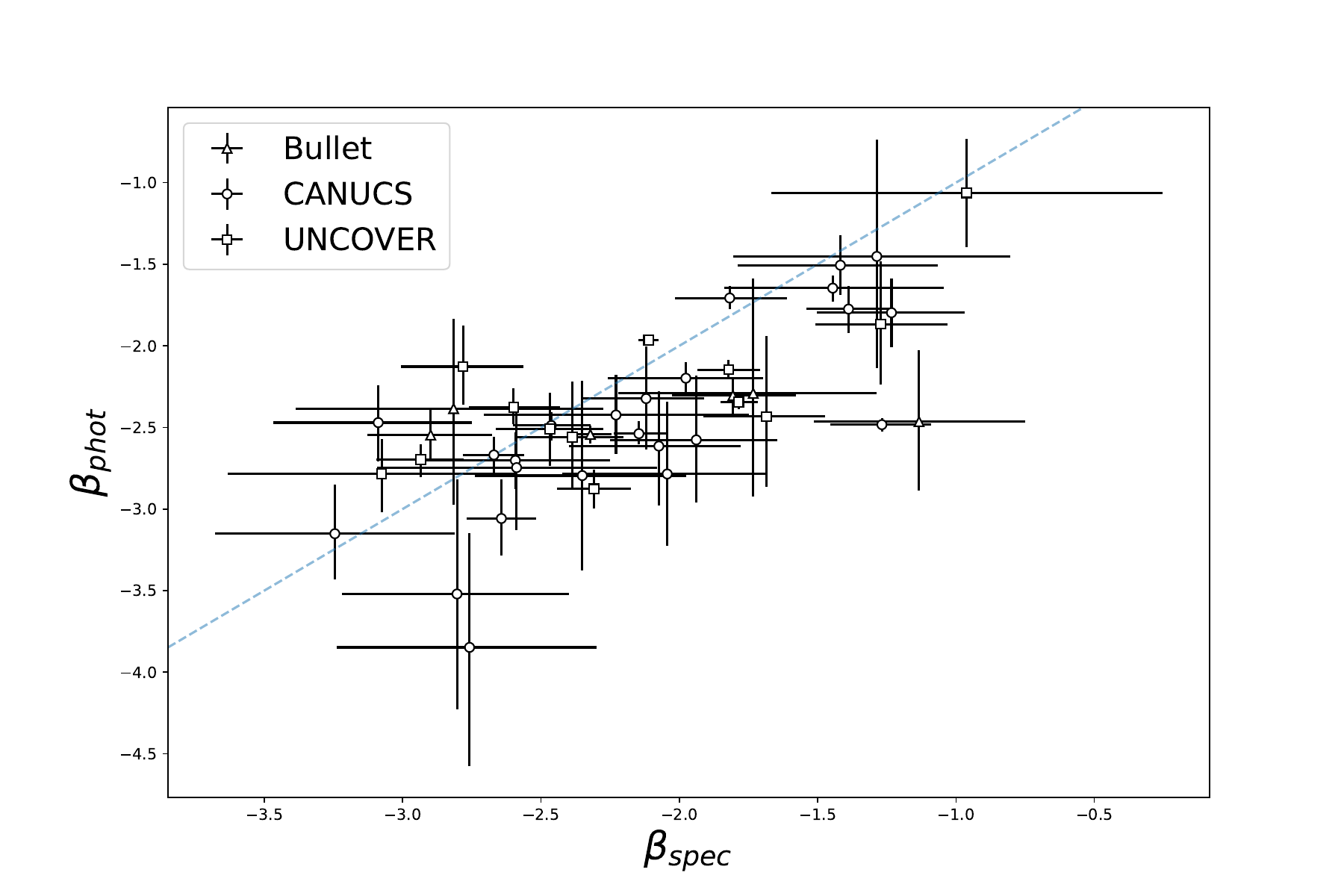}
    \caption{\textit Comparison between photometric and spectroscopic values for the UV slope $\beta$ index for 43 galaxies in our sample. The dashed blue line represents $\beta_{phot}=\beta_{spec}$.}
    \label{fig:betaphot-vs-betaspec}
\end{figure}

We subsequently compared the photometric ($\beta_{phot}$) and $\beta_{spec}$ for these objects, finding an average value $\Delta\beta = \beta_{phot}-\beta_{spec}=-0.08\pm 0.48$ (see \autoref{fig:betaphot-vs-betaspec}). This shows that the values for the UV slope calculated from photometry are consistent with spectroscopy on average. 
\subsection{UV magnitudes}
We estimated the absolute UV magnitude $M_{UV}$ for each galaxy in all cluster fields  using the brightness $F_\nu$ in the NIRCam filter closest to 1500$\AA$, which does not include the Ly$\alpha$ line. The brightness was corrected with the estimated lensing magnification $\mu$ of each object. We then rescaled the magnitude found with this filter to $1500\AA$ with the following formula:
$$M_{1500\AA}=M_{\lambda_{\text{pivot}}}-2.5(\beta+2)\text{log}_{10}(\frac{1500\AA}{\lambda_{\text{pivot}}})$$ 
where $\lambda_\text{pivot}$ is the pivot wavelength of the filter used. The uncertainties on $F_\nu$,$\mu$, and $\beta$ were then propagated in order to obtain the uncertainty on $M_{UV}$ \par
The best-fit values of $\mu$ were obtained from the lensing model for MACS0416 \citep{rihtarsic_canucs_2025}, Abell 370 \citep{gledhill_canucs_2024}, MACS0417 (Desprez et al. \textit{in prep.}), MACS1423 (Desprez et al. \textit{in prep.}), MACS1149, and the Bullet cluster \citep{rihtarsic_mapping_2026}. The lensing models are also available in the CANUCS Data Release 1 (DR1) and described in \citet{sarrouh_canucstechnicolor_2026}. The lensing model for Abell 2744 is described in \citet{furtak_uncovering_2023} and \citet{price_uncover_2025}.\par

\section{Results}\label{sec:results}
\begin{figure}[h]
    \centering
    \includegraphics[width=1\linewidth]{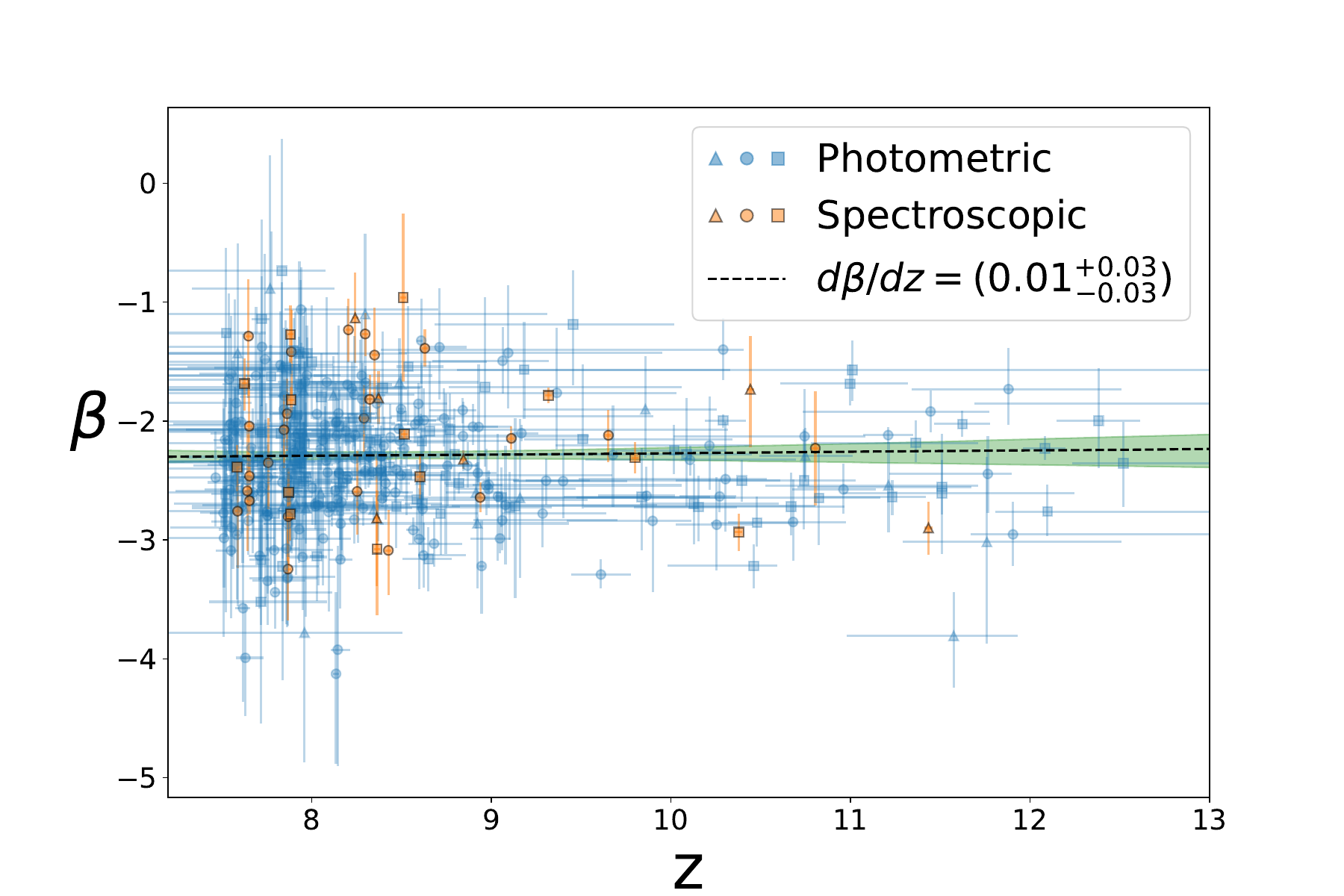}
    \caption{$\beta-z$ diagram for our high-redshift sample. Objects for which only photometric estimates of $\beta$ are available are plotted in blue, and the spectral estimates of $\beta$ are shown in orange. For objects that have both, we only plot the spectral estimates. The round, triangular, and square data points represent objects in the CANUCS, Bullet cluster, and UNCOVER samples, respectively. A linear fit of the trend and its 1$\sigma$ confidence level, obtained via a bootstrap method, are  represented by the dashed black line and the green shaded band.}
    \label{fig:beta-z}
\end{figure}

For our whole sample, we have an average value of $\beta=-2.1\pm0.4$. We weighted each value of $\beta$ with its inverse variance. The values range from -4.0 to -0.5 (see Fig. \ref{fig:beta-z}). Our objects also lie at an average redshift of $z=8.5\pm1.1$ and have an average $M_{UV}=-18.9\pm1.1$.\\
In \autoref{fig:beta-z} we show the trend between the UV slope and redshift of all galaxies in our sample. Fitting the whole sample with a total least-squares (TLS) regression by taking uncertainties on $\beta$ and redshift into account, we found no significant trend between $\beta$ and redshift: $d\beta/dz = 0.01\pm0.03$. We measured the uncertainty of this relation by bootstrapping the whole sample 1000 times and taking the 16th and 84th percentiles as lower and upper boundaries.

\begin{figure}[h]
    \centering
    \includegraphics[width=1\linewidth]{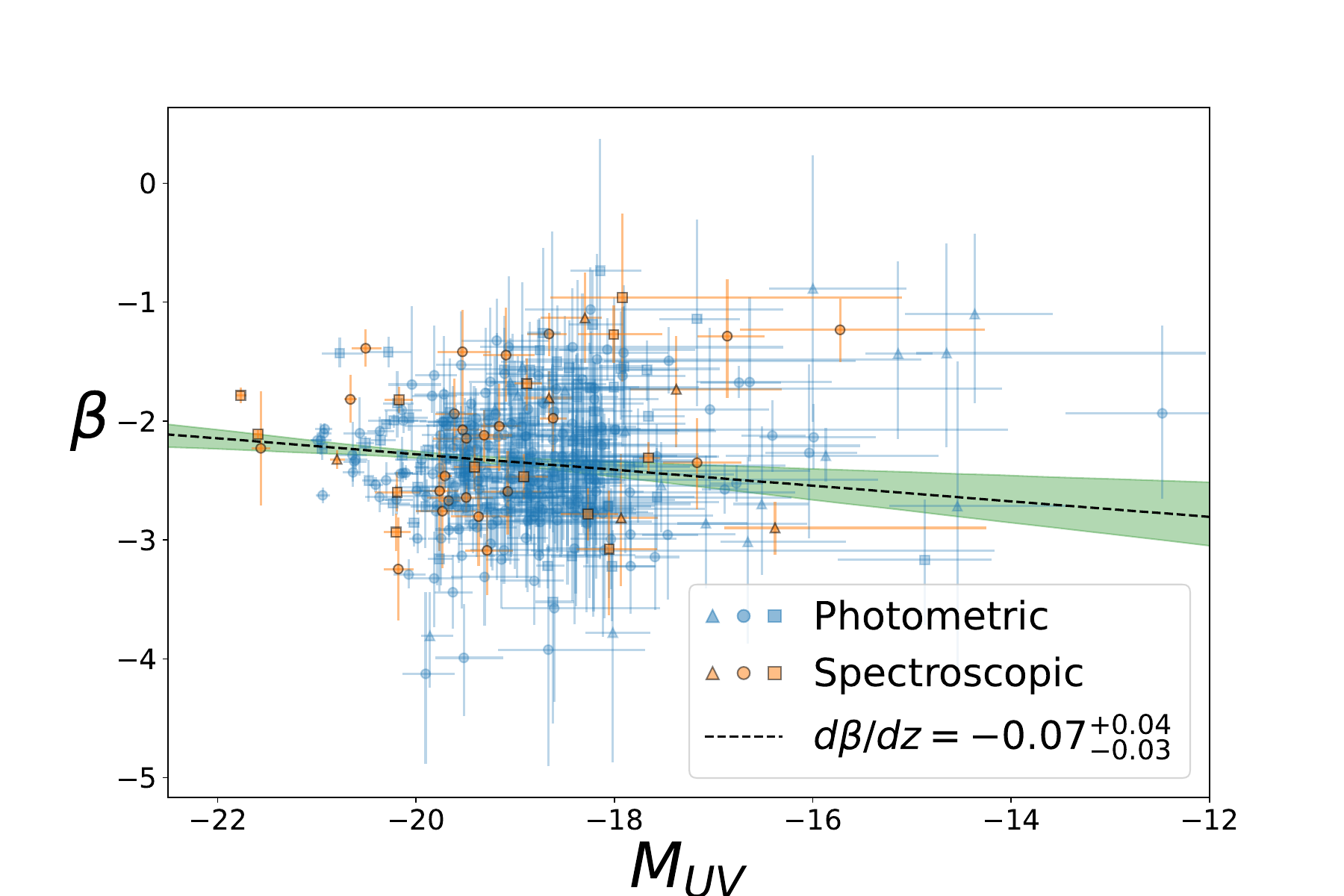}
    \caption{$\beta$-$M_{UV}$ diagram. The colour code is the same as in \autoref{fig:beta-z}. The dashed black line represents the relation found between $\beta$ and $M_{UV}$ for the whole sample, and the green shaded region represents a 1$\sigma$ confidence band.}
    \label{fig:beta-muv-all}
\end{figure}
\begin{figure}
    \centering
    \includegraphics[width=1\linewidth]{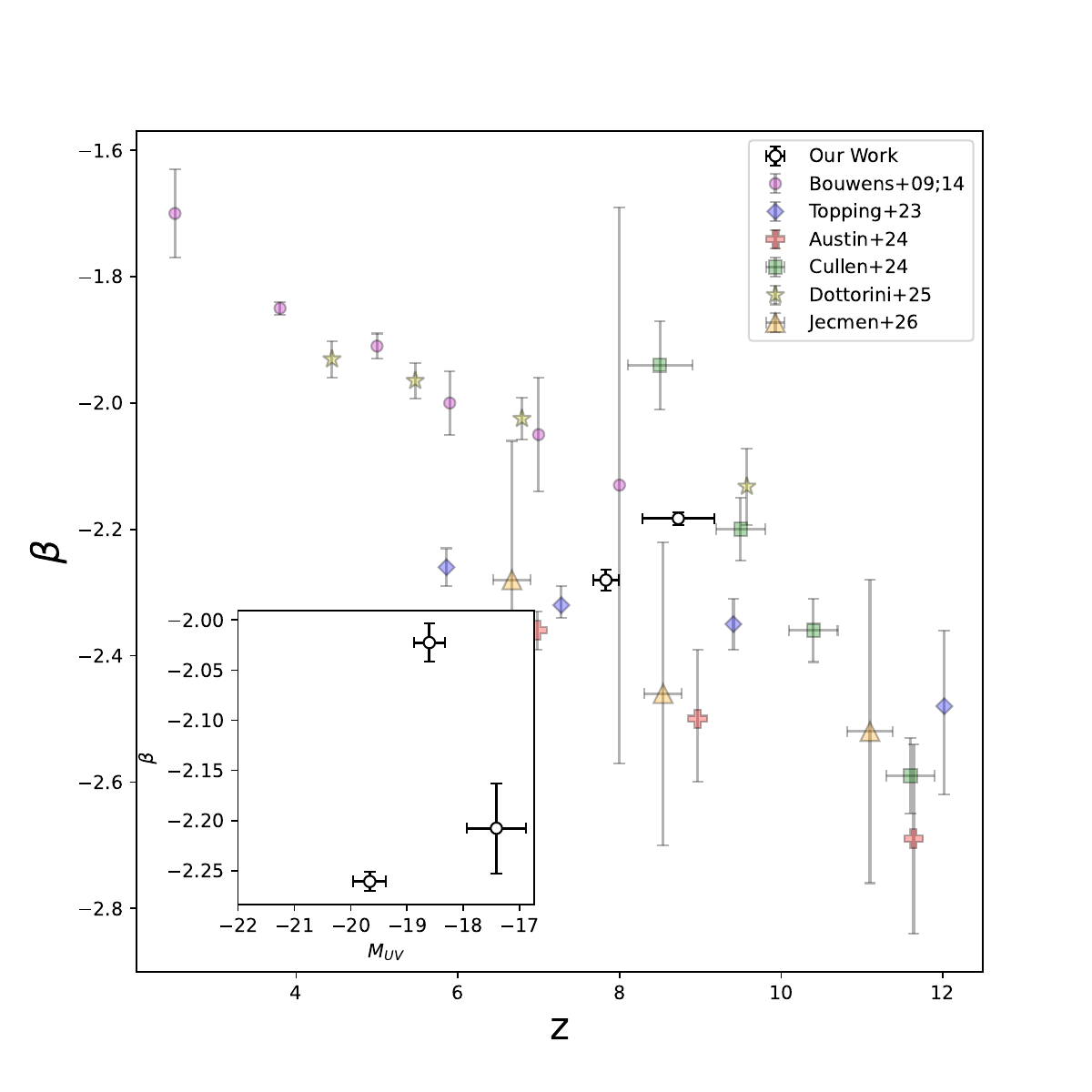}
    \caption{$\beta$-$z$ diagram (\textit{inset}: $\beta-M_{UV}$ diagram) for our sample in redshift ($M_{UV}$) bins. Our data is represented as white circles, reporting in the vertical direction the weighted mean of $\beta$ in each bin, with the standard deviation of each subsample as its error bar, and in the horizontal direction reporting the median and median absolute deviation ($M_{UV}$) in each bin. In coloured data points, values from previous works: purple circles from \citet{bouwens_uv-continuum_2014}, blue diamonds from \citet{topping_uv_2023}, red crosses from \citet{austin_epochs_2025}, green squares from \citet{cullen_ultraviolet_2024}, yellow stars from \citet{dottorini_evolution_2024}, and orange triangles from \citet{jecmen_glimpse_2026}.}
    \label{fig:beta_z_MUV_bins_weighted}
\end{figure}

In \autoref{fig:beta-muv-all} we show the relation between $\beta$ and $M_{UV}$, colour-coded as in \autoref{fig:beta-z}. We find $d\beta/dM_{UV} = -0.07^{+0.04}_{-0.03}$ throughout our whole sample, suggesting a very weak trend of the UV slope with galaxy brightness, at least compared to pre-JWST results (see \citealt{bouwens_uv-continuum_2014}, who found a slope of $-0.20\pm0.07$ for their sample at $z\sim7$). This agrees with the flatter slope reported in the more recent literature, however (e.g. \citealt{saxena_hitting_2024} with $-0.07$). \\
In \autoref{fig:beta_z_MUV_bins_weighted} we binned the data to compare our results with the literature, and we find similar results. However, due to the large scatter in our data and the limited statistics, our results are very sensitive to the choice of binning.

\subsection{LyC escape fraction}\label{sec:f_esc}

Several works recently tried to estimate the escape fraction of LyC photons in high-redshift galaxies, either from relations calibrated at low-redshift (\citealp{chisholm_far-ultraviolet_2022, lin_empirical_2024}) or using predictions from cosmological simulations (\citealp[e.g.][]{saxena_hitting_2024, choustikov_physics_2024}). Some models did not rely on low-redshift calibration and instead used an SED fitting of the spectral features to model the effect of LyC leakage on the emission lines and UV slope (see e.g. \citealp{giovinazzo_breaking_2025}). We used the model presented in \citet{mascia_new_2024}, which was calibrated on the low-redshift galaxy sample of \citet{flury_low-redshift_2022}. It uses the half-light radius $r_e$, the equivalent width of H$\beta$ $EW_0(H\beta)$, and the $\beta$ index of the UV slope of a galaxy to predict its LyC escape fraction,\par
$$\text{log}_{10}\text{f}_{esc} = A + B\cdot EW(H\beta)+ C \cdot r_e + D\cdot\beta$$.
The values of the coefficients A, B, C, and D are those reported in \citet{mascia_new_2024}.\par
In our sample, it was possible to obtain measurements for all three properties for 14 galaxies in total, as $H\beta$ measurements require NIRSpec data. The size measurements were obtained by Judež et al. (\textit{in prep.}) with the same method as in \citet{willott_search_2025}: using a forward-modeling procedure of galaxy shapes through the strong-lensing field with the \textsc{Lenstruction} software \citep{yang_versatile_2020, birrer_gravitational_2015, birrer_lenstronomy_2018}.\\
For one of these objects, the extrapolation yielded an unphysical value of $f_{esc}>1$ because of an extremely high $H\beta$ equivalent width of $(886\pm41)\AA$.
This object was not included in the final $f_{esc}$ sample, which finally consisted of 14 galaxies.\par
The confidence intervals were obtained with an MC method by resampling the value of each parameter and coefficient 1000 times within its distribution. The detailed results are reported in \autoref{tab:f_esc}. We find an average escape fraction of $0.26\pm0.22$ in our sample. This is slightly higher than the averages found in \citet{mascia_new_2024} for objects at $<z>=8$, but the dispersion is far higher.

\begin{table*}[t]
\centering
\caption{Physical properties of the 14 objects in our $f_{esc}$ sample}
\begin{tabularx}{\textwidth}{r r r X X X X}
\hline
CANUCS ID & RA & DEC & $EW_0(H\beta)$ & $r_e$ (kpc) & $\beta$ & $f_{esc}$(LyC) \\
\hline
1106089 & 64.376136 & -11.908745 & 33 $\pm$ 9 & 0.02$_{-0.01}^{+0.01}$ & -3.0 $\pm$ 0.3 & 0.20$_{-0.07}^{+0.30}$ \\
2110006 & 39.975840 & -1.587273 & 190 $\pm$ 40 & 0.110$_{-0.005}^{+0.005}$ & -2.6 $\pm$ 0.3 & 0.30$_{-0.10}^{+0.50}$ \\
2120090 & 39.966109 & -1.594787 & 440 $\pm$ 70 & 0.045$_{-0.019}^{+0.012}$ & -1.8 $\pm$ 0.5 & 0.8$_{-0.6}^{+2.3}$ \\
3107165 & 64.039230 & -24.093198 & 170 $\pm$ 30 & 0.54$_{-0.01}^{+0.01}$ & -1.7 $\pm$ 0.1 & 0.05$_{-0.01}^{+0.03}$ \\
3113053 & 64.043179 & -24.057928 & 70 $\pm$ 20 & 0.034$_{-0.003}^{+0.006}$ & -3.2 $\pm$ 0.4 & 0.30$_{-0.10}^{+0.70}$ \\
3114538 & 64.039317 & -24.045717 & 10 $\pm$ 10 & 0.16$_{-0.02}^{+0.02}$ & -2.4 $\pm$ 0.3 & 0.085$_{-0.027}^{+0.092}$ \\
3117269 & 64.048130 & -24.081449 & 110 $\pm$ 20 & 0.269$_{-0.009}^{+0.01}$ & -1.6 $\pm$ 0.1 & 0.061$_{-0.019}^{+0.022}$ \\
4100225 & 215.942388 & 24.069652 & 210 $\pm$ 140 & 0.13$_{-0.009}^{+0.01}$ & -2.5 $\pm$ 0.1 & 0.38$_{-0.09}^{+0.70}$ \\
4117337 & 215.944441 & 24.068745 & 150 $\pm$ 180 & 0.022$_{-0.003}^{+0.004}$ & -1.7 $\pm$ 0.2 & 0.12$_{-0.033}^{+0.27}$ \\
4117358 & 215.941214 & 24.069316 & 20$\pm$ 180 & 0.147$_{-0.007}^{+0.007}$ & -2.2 $\pm$ 0.1 & 0.071$_{-0.017}^{+0.14}$ \\
5101296 & 177.379426 & 22.345013 & 330 $\pm$ 10 & 0.056$_{-0.025}^{+0.031}$ & -2.3 $\pm$ 0.5 & 0.63$_{-0.16}^{+0.94}$ \\
5104911 & 177.359469 & 22.375533 & 100 $\pm$ 100 & 0.045$_{-0.019}^{+0.024}$ & -2.0 $\pm$ 0.3 & 0.13$_{-0.036}^{+0.35}$ \\
5110240 & 177.389947 & 22.412704 & 220 $\pm$ 80 & 0.087$_{-0.007}^{+0.009}$ & -2.5 $\pm$ 0.2 & 0.36$_{-0.10}^{+0.53}$ \\
5112687 & 177.390930 & 22.349767 & 90 $\pm$ 80 & 0.029$_{-0.013}^{+0.015}$ & -1.8 $\pm$ 0.3 & 0.098$_{-0.029}^{+0.21}$ \\
\hline
\end{tabularx}

\label{tab:f_esc}
\end{table*}

\section{Discussion}\label{sec:discussion}
\subsection{Sample heterogeneity}
While the sample might be heterogeneous because it was assembled from fields with a different photometric coverage, we note that the Bullet cluster sample was observed at the same depth and with the same filter set as the CLU fields in CANUCS. In addition, the medium-band filters do not alter the measurement of $\beta$ between CLU and NCF fields significant. We tested this by measuring the UV slopes in the NCF fields with the same set of filters as the CLU pointings, and the $\beta$ we measured were within the uncertainty bounds. Finally, the UNCOVER and MEGASCIENCE survey covers Abell 2744 with the same set of filters as CANUCS and Technicolor in three of the CANUCS fields, and with similar depths. Our results are not changed when the same analysis is conducted without the sources in Abell 2744.
\subsection{UV slope trends}
The absence of a trend of $\beta$ with redshift contradicts previous studies at lower redshifts (e.g. \citealp{bouwens_uv-continuum_2014, bolamperti_uv-continuum_2023}), which all reported a negative trend with bluer slopes at higher redshifts. However, more recent studies made possible by JWST included galaxies at redshifts $z>7$ also found a flattening of this relation at the highest redshifts (see e.g. \citealt{saxena_hitting_2024, roberts-borsani_jwst_2025, jecmen_glimpse_2026, austin_epochs_2025}). Our results apparently support these latest studies and confirm the flattening of the overall slope in the UV at the higher redshifts. The flattening of this relation might indicate that the UV emission of more galaxies at high-z is dominated by the nebular continuum with an intrinsic $\beta\sim-2.3$. This might explain the lack of a continuing trend at lower redshift. On the other hand, we also note the limited statistics of our sample, especially at $z>9$, and the redshift ranges $7.5<z<9$ and $9<z<12$ only span 150 to 200 Myr each, which may not be enough to register a change in the overall UV slope.\par 
We also found a very weak trend between $\beta$ and $M_{UV}$, with lower $\beta$ for fainter galaxies. This contradicts studies at lower redshift, which usually reported that fainter and less massive galaxies have bluer slopes \citet{bouwens_very_2010, bouwens_uv-continuum_2014, topping_uv_2023}. We instead found a weak relation, which agrees with what was found by \citet{saxena_hitting_2024}, who found a similar trend for a subsample at $z>8$. This seems to give credit to the model in which high- and low-mass galaxies achieve similarly steep slopes during the first star-forming episodes, but more massive and brighter galaxies accumulate dust more rapidly, leading to redder UV slopes at lower redshift. The trend between UV slope and magnitude is affected by a large scatter in the galaxy population, however, which is compounded by large uncertainties in lensing-model magnifications. This makes this relation less robust. \par
\subsection{Differences in the photometric and spectroscopic measurement of $\beta$}
We found consistent measurements between photometric and spectroscopic UV slopes in our sample, with $<\Delta\beta>=0.2\pm0.4$. The offset is significant for a handful of objects, however. For a few of these, the discrepancy is driven by the different regions of a galaxy that are captured by spectroscopic and photometric procedures. For the Firefly Sparkle galaxy \citep{mowla_formation_2024}, for example, the spectrum was mainly obtained from the central clump, while the $0^{\arcsec}_. 3$ aperture photometry also captures the light from two adjacent much bluer clumps, and therefore, bluer slopes are measured from the photometry. In other cases, the discrepancy is due to the number of photometric filters used to measure the UV slope and sampling part of the spectrum redward of the \citet{calzetti_dust_1994} windows. This might lead to a discrepancy especially for the galaxies in the CLU field, for which fewer filters are available, only two per galaxy in several cases, in order to measure the slope. This issue is alleviated in the NCF fields, where medium-band filters are available in the same wavelength range in addition to the broad-band filters.\par
\subsection{Extrapolation of LyC $f_{esc}$}
Finally, we extrapolated the values of the LyC escape fraction $f_{esc}$ for a subsample of 14 galaxies observed with NIRSpec and for which we have emission line measurements (see \autoref{tab:f_esc}).  We found an average escape fraction of $0.26\pm0.22$ throughout the subsample. This value is moderately high with respect to the $\sim10\%$ that are required to ionise most of the IGM by $z\sim6$ \citep{finkelstein_conditions_2019}, but the wide scatter means that it agrees with values found by \citet{mascia_new_2024}. Our values should be taken with caution because the data that were used to calibrate the relation in \citet{mascia_new_2024} were taken from a sample of galaxies at low redshift \citep{flury_low-redshift_2022} where the direct measurement of $f_{esc}$ is possible, and our extrapolation assumed that the same calibration would hold true at the EoR. Other recent studies also relied on local multivariate calibrations (e.g. that of \citealt{jaskot_multivariate_2024}) found relatively high values of the escape fraction: \citep{jecmen_glimpse_2026} reported an average $f_{esc}=16\%$ in GLIMPSE, with the fraction peaking in galaxies at $M_{UV}\sim16.5$ and declining for fainter galaxies. The use of the low-z calibration to determine the escape fraction for high-z galaxies presents several issues. Local calibration relies on ionisation states, metallicity, and nebular continuum that differ in high-redshift galaxies, and these calibrations tend to break down at higher redshift \citep{laseter_jades_2024}. Crucially, as pointed out by \citet{jaskot_multivariate_2024}, high-redshift galaxies are more compact and have a smaller half-light radius $r_e$ than the galaxies in the LxLCS, and using $r_e$ means extrapolating outside of the parameter space used in the calibration. \citet{giovinazzo_breaking_2025} argued that thesee relations already no longer hold between $z\sim0.3$ and $z\sim3$ because the change in the galactic environment can affect these relations. At high redshift, the higher rate of galaxy mergers and other galaxy interactions would affect the star formation process and the creation or obscuration of LyC leakage channels. Recent results tended to find a relatively high ionising efficiency $\xi_{ion}$ \citep{harshan_canucs_2024, austin_resolving_2025} for galaxies at high redshift, which, combined with an $f_{esc}\sim0.26$, would lead to a difference in the ionising photon budget \citep{munoz_reionization_2024}. This might be lessened by considering a clumpy IGM \citep{davies_predicament_2024}, quickening hydrogen recombination, and increasing the needed budget of ionising photons in order to achieve reionisation by $z\sim6$ \citep{austin_resolving_2025}. More recent works that did not rely on local calibrations but instead used indirect measurements appeared to agree on a lower value for $f_{esc}$ of about $0.05\sim0.10$ \citep{papovich_galaxies_2026, giovinazzo_breaking_2025, austin_resolving_2025}. The extrapolations reported in \autoref{tab:f_esc} are therefore better treated as estimates that strongly dependent on local galaxy models and not as robust estimates of the escape fraction of a true high-redshift population.

\section{Summary and conclusions}\label{sec:summary}
We used photometric and spectroscopic data from the JWST NIRCam and NIRSpec instrument gathered by six surveys in seven strong-lensing fields: CANUCS for the clusters MACS0417, MACS0416, Abell370, MACS1149 and MACS1423, Technicolor and JUMPS for MACS0416, Abell370, and MACS1149,  \textit{Silver Bullet for Dark Matter} for the Bullet cluster, and GLASS, UNCOVER and MEGASCIENCE for Abell 2744. We analysed the UV slopes ($\beta$) in a sample of 378 galaxies at $7.5<z<12.5$. The high magnification provided by the seven cluster fields targeted by these surveys enabled us to recover galaxies down to a magnitude of $M_{UV}\lesssim-16$ in the main cluster pointing.  We were also able to more robustly estimate values of $\beta$ in the NCF fields based on the numerous medium-band filters employed by CANUCS, TECHNICOLOR, and UNCOVER. The combination of these two factors, the very deep data in the CLU pointing, and the very accurate data in the NCF allowed us to analyse the trends of $\beta$ reliably. We compared our results with the spectroscopic $\beta$ slopes measured for 43 of the galaxies after correcting the spectra with photometry, and we found no statistically significant offset of bluer slopes with photometry with respect to spectroscopy. We found a few individual objects with significant offsets, however, mainly because different parts of each galaxy are probed by NIRCam and NIRSpec, while the effect of possible slit losses on the UV slope when observing with the NIRSpec instrument was addressed by correcting the spectrum with the photometry itself.
Our results are listed below.
\begin{enumerate}

    \item There is no significant trend between $\beta$ and redshift, with $d\beta/dz = 0.01^{+0.03}_{-0.03}$ over the range $7.5<z<12.5$, consistent with a flattening of the relation past $z>7$ found in recent works. This underlines the similarity in average galaxy properties between $\sim300$ and $\sim600$ Myr after the Big Bang, with most of the scatter due to the variance in galaxy properties, and possibly due in part to a strong contribution from the nebular continuum and an absence of dust, which would redden the slopes.
    \item There is a weak trend between $\beta$ and $M_{UV}$ $d\beta/dM_{UV}=-0.07^{+0.04}_{-0.03}$ over the whole UV magnitude range of $-22<M_{UV}<-16$, also consistent with recent results at high redshift (\citealp{saxena_hitting_2024}, \citealp{austin_epochs_2025}). This might indicate a lack of dust, or at least an absence of a correlation between dust and SFR for galaxies in the interval $7.5<z<12.5$, with more massive galaxies not having had enough time yet to build up a dust reservoir with respect to less massive galaxies.
\end{enumerate}
\begin{acknowledgements}
GF, MB, JJ, GR, NM, AH, and VM acknowledge support from the ERC Grant FIRSTLIGHT \# 101053208, Slovenian national research agency ARIS through grants N1-0238 and P1-0188, and ESA PRODEX Experiment Arrangements No. 4000146646 and 4000149972. This research was also enabled by grant 18JWST-GTO1 from the Canadian Space Agency and funding from the Natural Sciences and Engineering Research Council of Canada. This research used the Canadian Advanced Network For Astronomy Research (CANFAR) operated in partnership by the Canadian Astronomy Data Centre and The Digital Research Alliance of Canada with support from the National Research Council of Canada the Canadian Space Agency, CANARIE and the Canadian Foundation for Innovation. The data were obtained from the Mikulski Archive for Space Telescopes at the Space Telescope Science Institute, which is operated by the Association of Universities for Research in Astronomy, Inc., under NASA contract NAS 5-03127 for JWST. Support for program \#4598 was provided by NASA through a grant from the Space Telescope Science Institute, which is operated by the Association of Universities for Research in Astronomy, Inc., under NASA contract NAS 5-03127. Some of the data products presented herein were retrieved from the Dawn JWST Archive (DJA). DJA is an initiative of the Cosmic Dawn Center (DAWN), which is funded by the Danish National Research Foundation under grant DNRF140.
\end{acknowledgements}

\bibliographystyle{aa}
\bibliography{references}
\end{document}